\documentclass[a4paper]{jpconf}
\usepackage{amsmath,amsfonts,amssymb,wasysym}
\usepackage{pgf}
\usepackage{xcolor}
\usepackage{fancyvrb}
\usepackage{comment}
 \usepackage{sidecap}
\usepackage{graphicx}
\usepackage{framed}
\usepackage{epstopdf}
\usepackage{mathtools}
\usepackage{mathrsfs}
\usepackage{tikz}
\usepackage[absolute,overlay]{textpos}
\usetikzlibrary{shapes,snakes}
\usepackage{multirow}
\usepackage{changepage}
\usepackage{hyperref}

\newcommand{\gev}{{\unskip\,\text{GeV}}}

\newcommand{\eq}[1]{Eq.~(\ref{#1})}

\newcommand{\fig}[1]{Fig.~\ref{#1}}
\newcommand{\tab}[1]{Table~\ref{#1}}

\newcommand{\sect}[1]{Section~\ref{#1}}

\newcommand{\be}{\begin{equation}}
\newcommand{\ee}{\end{equation}}
\newcommand{\bea}{\begin{eqnarray}}
\newcommand{\eea}{\end{eqnarray}}

\newcommand{\crn}{\nonumber \\}

\newcommand{\fr}{\frac}
\newcommand{\bc}{\begin{center}}
\newcommand{\ec}{\end{center}}

\newcommand{\varep}{\varepsilon}

\newcommand {\ba}{\begin{array}}
\newcommand {\ea}{\end{array}}
\newcommand{\ben}{\begin{enumerate}}
\newcommand{\een}{\end{enumerate}}

\begin{document}
\title{Effects of bottom quark induced processes on polarized $W^+W^-$ production at the LHC at NLO}
\author{Duc Ninh Le, Thi Nhung Dao}
\address{Faculty of Fundamental Sciences, PHENIKAA University, Hanoi 12116, Vietnam}
\ead{ninh.leduc@phenikaa-uni.edu.vn}
\begin{abstract}
In this report we discuss the definition of the polarized 
cross sections of the inclusive $W^+W^-$ production at the LHC. 
Results at the level of next-to-leading order (NLO) QCD+EW accuracy, published in our recent paper, 
are presented to highlight the effects of bottom-quark induced processes. 
Compared to the unpolarized case, the bottom-induced effects after the subtraction of 
the on-shell top-quark contribution are more sizable for the doubly-longitudinal polarization. 
\end{abstract}
\section{Introduction}
\label{sec:into}
Proton-proton collisions at the LHC provide us valuable information into the world of fundamental 
particles. A lot of data have been collected and stored, and more are being produced with the 
current LHC run-3, which is scheduled until June 2026. 
The current colliding energy of the LHC is $13.6$ TeV, increased 
from the $13$ TeV value of run-2. 
These data can be used to explore the limits of the Standard Model (SM), which is 
based on the local gauge symmetry of $SU(3)_C \otimes SU(2)_L \otimes U(1)_Y$ 
and the electroweak spontaneous symmetry breaking mechanism. 

Among many different kinds of measurements, polarization measurements of massive gauge boson 
pair production at the LHC have recently attracted a significant amount of attention. 
ATLAS and CMS have published results of polarized diboson production in 
\cite{ATLAS:2022oge,ATLAS:2024qbd} ($WZ$) and 
\cite{ATLAS:2023zrv} ($ZZ$), and \cite{CMS:2020etf} (same-sign $WWjj$).
These efforts help us to explore unknown regions of particle physics, including: 
(i) the on-shell and off-shell boundary, 
(ii) different polarization modes and their interference. 

Differently from the unpolarized case, polarization measurements must somehow 
separate the polarized cross sections from the unpolarized cross section. 
As the detectors can only provide information of the decay products, 
an additional step must be done to define polarizations 
in the measurement. 
This definition must give a concrete result for kinematic distributions of the 
decay products. 
This set of polarized distributions (named polarization templates), which can only be produced by simulation (e.g. using the SM), 
is the definition of the polarizations. 
If all unpolarized differential cross section measurements can be fitted by 
a sum of those polarization templates (including a template for their interference), then we say that 
the polarization model is correct and the polarization fractions ($\sigma_\text{pol-i}/\sigma_\text{unpol}$) 
can be extracted from a fit to a chosen distribution 
(in principle, ``any'' kinematic distribution can be chosen, but some give better results than others 
due to different discrimination power,
e.g. a boosted-decision-tree score distribution was chosen in \cite{ATLAS:2023zrv}. 
This step can be done for the whole fiducial phase space or bin-wise.). 
In reality, off-shell effects are included in the unpolarized cross section measurements. 
Therefore, these effects must be accounted for by using a separate template or 
including them in the background as done in \cite{ATLAS:2023zrv}.   

It then becomes clear that polarization templates are one of the most important information in a polarization 
measurement. Ideally, these templates should be published together with the paper, or at least 
a description of them should be provided. 

It is also clear that polarization templates depend on the theoretical model used in the 
simulation and the precision of the cross section calculation. 
The precision level of the simulation changes with time. 
This is of course true for any measurements (e.g. the background estimation using simulation changes with new updates), 
but the impact on polarization measurements 
may be greater than on the unpolarized cases. 

As parton showers are expected to be different for different polarized $VV$ modes, 
these effects have to be calculated separately for each polarization template. 
Work in this direction 
has been done, see \cite{Hoppe:2023uux, Pelliccioli:2023zpd, Javurkova:2024bwa}. 
Hadronization is also different for different polarizations, this effect is however 
expected to be much smaller than those of parton showers. 

Kinematic cuts affect polarization states differently. 
To allow for future (re-)analyses with different lepton kinematic 
cuts, a good strategy is to 
generate events for different polarization states with an inclusive cut setup 
(e.g. including only $m_{4l} > M_V + M_{V^\prime}$ cut)
and store these events. 
At the analysis level, analysis cuts are then applied to these inclusive polarized 
events to produce fiducial-level templates to be used in the final fit. 
The extracted polarization fractions depend on the analysis cuts 
as the cuts affect $\sigma_\text{pol-i}$ and $\sigma_\text{unpol}$ 
differently.

Since parton showers can only improve results in the soft and collinear regions, 
fixed order calculations are needed for a better description of the hard regions. 
Next-to-leading order (NLO) and beyond calculations of the polarized cross sections are therefore important. 
Progress has been made in this direction as well. 

We mention here the recent development in the calculation of higher-order corrections to 
the polarized $VV$ production cross sections with fully leptonic decays. 
The first NLO QCD corrections were done in \cite{Denner:2020bcz} for the $W^+W^-$ case. 
These results were soon extended to the $ZZ$ case \cite{Denner:2021csi}, including the NLO electroweak (EW) corrections this time. 
This $ZZ$ work is important as it provides new on-shell mappings (see \sect{sect:cal}) $1\to 2$ and 
$1\to 3$, which are crucial for the definition of the polarization templates. 
The $1\to 2$ mapping is used for the leading order (LO) $V$ decays, while the 
$1\to 3$ mapping is for the NLO $V$ decays occurring in the NLO EW corrections.
Using these new on-shell mappings, the NLO EW corrections for the $WZ$ case 
were completed in \cite{Le:2022lrp,Le:2022ppa}. 
This is a major improvement compared to the $ZZ$ case because the $WZ$ production 
involves an on-shell charged current where a photon can be radiated off the 
intermediate on-shell $W$. This radiation introduces soft divergences which 
must be canceled with the counter parts in the virtual corrections of 
the $WZ$ production and $W$ decay amplitudes. 
The $WZ$ calculation was then straightforwardly extended to the $W^+W^-$ case 
in \cite{Denner:2023ehn,Dao:2023kwc}. 
A further step was recently done for the same-sign $W^+W^+jj$ production \cite{Denner:2024tlu}, where 
final-state photon radiation is more complicated than the inclusive $W^+W^-$ case 
as interferences between jet-photon radiation amplitudes and 
lepton-photon radiation amplitudes occur. Treatment of 
these radiations requires modifications of the decay momenta. 
These modifications interplay with the on-shell mappings, therefore require careful attention. 
Moreover, the important next-to-next-to-leading order (NNLO) QCD corrections for the polarized $W^+W^-$ production were 
calculated in \cite{Poncelet:2021jmj}.
The case of semi-leptonic decays has been considered in \cite{Denner:2022riz}. 

In this report, which is a contribution to the proceedings of the 
{\tt PIAS workshop 2024: Physics at different scales} (Hanoi, Vietnam, 2024), 
we consider the case of inclusive $W^+W^-$ production and focus on 
the definition of the polarized $W^+W^-$ pair production. This is the signal part. 
The definition used in this work is in accordance with the common definition of the 
unpolarized $W^+W^-$ pair production used in the literature and in the experimental measurements. 
This definition requires the subtraction of the top-quark contribution. 
This subtraction has to be done for individual polarizations and 
has been recently achieved by us in \cite{Dao:2024ffg} at NLO QCD+EW accuracy. 
The presentation presented here is based on this result. 
In addition, we take this opportunity to discuss in more detail the 
definition of the inclusive polarized $W^+W^-$ production signal.  
This discussion is provided in \sect{sect:cal}. 
Important numerical results are then presented in \sect{sect:res}. 
Finally, a short conclusion is provided in \sect{sect:con}.
\section{Method of calculation}
\label{sect:cal} 
\begin{figure}[h]
  \centering
  \includegraphics[width=0.6\textwidth]{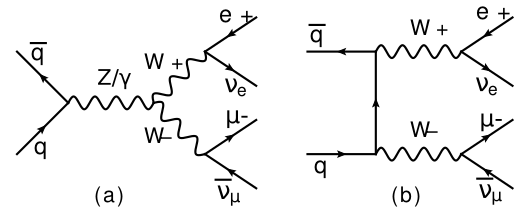}
  \caption{Tree-level Feynman diagrams describing a $W^+W^-$ resonant signal process at the LHC.}
  \label{diag_WW_signal}
\end{figure}

The signal process is
\bea
p(k_1) + p(k_2) \to W^+(q_1) W^-(q_2)
\to e^{+}(k_3) + \nu_e(k_4) + \mu^{-}
(k_5) + \bar{\nu}_\mu(k_6) ,
\label{eq:proc1}
\eea
where the final-state leptons are the decay product of the intermediate $W^+W^-$ system. 
This is illustrated by a Feynman diagram in \fig{diag_WW_signal}.

A $W$ boson can decay into leptons ($33\%$) or quarks ($67\%$). 
The hadronic final state has a higher rate, but suffers from a much larger background than 
the leptonic mode. We therefore only consider the fully leptonic decays of the $W$ bosons in this work. 

The two intermediate $W$ bosons are required to be on-shell, namely their momenta satisfying 
$q_i^2 = M_W^2$ with $i=1,2$. This condition is crucial for the separation of the different 
polarization contributions, because gauge invariance is guaranteed in this on-shell limit. 

To make this on-shell phase space available, the invariant mass of the final-state leptons 
must be greater than $2M_W$. It is therefore important that this condition is also satisfied 
in the experimental analysis. 

In simulation, the different polarizations of the $W^+W^-$ pair 
can be separated using the double-pole approximation (DPA). 
This technique has been used for the $ZZ$ \cite{Denner:2021csi}, $WZ$ \cite{Denner:2020eck,Le:2022lrp,Le:2022ppa}, and $W^+W^-$ \cite{Denner:2020bcz,Denner:2023ehn,Dao:2023kwc,Dao:2024ffg} processes 
up to NLO QCD+EW level. For the $W^+W^-$ case, an NNLO QCD calculation has 
been performed \cite{Poncelet:2021jmj}. 

The idea of the DPA can be understood by considering the unpolarized LO 
DPA amplitude: 
\bea
\mathcal{A}_\text{DPA}^{\bar{q}q\to V_1V_2\to 4l} = \fr{1}{Q_1Q_2}
\sum_{\lambda_1,\lambda_2=1}^{3}
\mathcal{A}_\text{LO}^{\bar{q}q\to V_1V_2}(\hat{k}_i,\lambda_1,\lambda_2)\mathcal{A}_\text{LO}^{V_1\to
    l_1l_2}(\hat{k}_i,\lambda_1)\mathcal{A}_\text{LO}^{V_2\to l_3l_4}(\hat{k}_i,\lambda_2)
,\label{eq:LO_DPA}
\eea
with 
\bea
Q_j = q_j^2 - M_{V_j}^2 + iM_{V_j}\Gamma_{V_j}\;\; (j=1,2),
\label{eq:Qi_def}
\eea
where $M_{V_j}$ and $\Gamma_{V_j}$ are the mass and the total decay width of the $W$ bosons.  

Compared to the full $\bar{q}q\to 4l$ amplitudes, contributions from the non-doubly-resonant 
diagrams have been omitted. Another notable difference is the application of the 
on-shell phase space mapping 
\bea
\text{on-shell mapping:}\;\;\; [k_i] \to [\hat{k}_i]
\eea
with $i=\overline{1,6}$ for all external-state particles. 
$k_i$ are the off-shell momenta. 
The mapped (or on-shell) momenta $\hat{k}_i$ are all massless and satisfy the on-shell condition:
\bea
\hat{q}_1^2 = (\hat{k}_3 + \hat{k}_4)^2 = M_W^2, \;\;\; 
\hat{q}_2^2 = (\hat{k}_5 + \hat{k}_6)^2 = M_W^2. 
\eea 
Solution for these on-shell momenta can be found if the off-shell momenta satisfy: 
\bea
(k_3 + k_4 + k_5 + k_6)^2 \ge 4M_W^2,
\eea
as above mentioned. 

The result of \eq{eq:LO_DPA} is obtained from the key observation that the on-shell production amplitudes 
$\mathcal{A}_{\bar{q}q\to V_1V_2}$ and on-shell decay amplitudes 
$\mathcal{A}_{V_i \to ll'}$ 
are individually gauge invariant. 
This is the reason for the application of the on-shell mapping. 
The DPA result of \eq{eq:LO_DPA} can be thought of as an on-shell projection 
of an off-shell amplitude. This on-shell projection includes the following steps:
\begin{itemize}
\item Select only the $V_1V_2$ doubly-resonant Feynman diagrams.
\item Replace the intermediate $V_i$ propagators using:
\begin{align}
\text{$V_i$ propagator:}\;\;\; P_i &= \fr{-ig^{\mu\nu}}{Q_i} \to 
\fr{i}{Q_i} \sum_{\lambda=1}^{3}\varep^{\mu *}_\lambda(q_i) \varep^\nu_\lambda(q_i),\\
g^{\mu\nu} &= -\sum_{\lambda=1}^{3}\varep^{\mu *}_\lambda(k) \varep^\nu_\lambda(k) + \fr{k^\mu k^\nu}{M_V^2},\;\; k = q_i, 
\label{eq:metric}
\end{align}
where the `t Hooft-Feynman gauge has been used for simplicity. We can choose another gauge which 
can introduce a new term proportional to $q_i^\mu q_i^\nu$ into $P_i$. 
This additional term cancels against the corresponding would-be Goldstone boson contribution. 
For the same reason, the last term proportional to $k^\mu k^\nu$ in \eq{eq:metric} 
can be safely removed. This cancellation is expected to occur at any order 
in a perturbation theory (see e.g. \cite{Denner:2024tlu}).  
\item Perform the on-shell momenta mapping for the production amplitudes and decay amplitudes.
\end{itemize}

\eq{eq:LO_DPA} provides a simple way to separate polarizations of the intermediate $W$ bosons. 
We just need to select the desired $(\lambda_1,\lambda_2)$ combination from the DPA amplitude. 
Specifically, we have
\begin{align}
\mathcal{A}_\text{LL} &= \mathcal{A}_\text{DPA}(\lambda_1=2,\lambda_2=2),\crn
\mathcal{A}_\text{LT} &= \mathcal{A}_\text{DPA}(\lambda_1=2,\lambda_2=1)+\mathcal{A}_\text{DPA}(\lambda_1=2,\lambda_2=3),\crn
\mathcal{A}_\text{TL} &= \mathcal{A}_\text{DPA}(\lambda_1=1,\lambda_2=2)+\mathcal{A}_\text{DPA}(\lambda_1=3,\lambda_2=2),\crn
\mathcal{A}_\text{TT} &= \mathcal{A}_\text{DPA}(\lambda_1=1,\lambda_2=1)+\mathcal{A}_\text{DPA}(\lambda_1=1,\lambda_2=3)\crn
&+\mathcal{A}_\text{DPA}(\lambda_1=3,\lambda_2=1)+\mathcal{A}_\text{DPA}(\lambda_1=3,\lambda_2=3).
\end{align}
From this we see that the LT contribution is a coherent sum of the $(2,1)$ and $(2,3)$ polarized amplitudes. 
Similarly can be said for the TL and TT polarizations. 
The unpolarized amplitude reads:
\begin{align}
\mathcal{A}_\text{unpol} = \mathcal{A}_\text{LL} + \mathcal{A}_\text{LT} + \mathcal{A}_\text{TL} + \mathcal{A}_\text{TT},
\end{align}
which gives
\begin{align}
\sigma_\text{unpol} = \sigma_\text{LL} + \sigma_\text{LT} + \sigma_\text{TL} + \sigma_\text{TT} + \sigma_\text{pol-int},
\end{align}
where the last term is called the polarization interference. 
This interference is the interference between the LL, LT, TL, and TT modes. 
This interference vanishes for the integrated cross section in the case of fully inclusive 
decay products, namely the momenta of the decay products are not restricted \cite{Denner:2020bcz}. 
If a particular distribution (e.g. $p_{T,\ell}$) is considered, then the interference effect 
will show up at a given bin, even for a fully inclusive analysis. 
This is because the cross section of a specific bin is no longer inclusive.  

The above on-shell projection has been extended up to NLO EW level 
in \cite{Denner:2021csi,Le:2022ppa,Denner:2023ehn,Dao:2023kwc,Dao:2024ffg} for the 
inclusive diboson case and in \cite{Denner:2024tlu} for the $W^+W^+jj$ case. 
The case of QCD corrections is simpler as only radiative corrections to the 
intitial-state particles are involved. NLO EW corrections are more complicated because 
radiative corrections occur for both initial and final state particles. 
Moreover, as the intermediate $W$ bosons are on-shell soft divergences occur 
when a real or virtual photon is emitted from these $W$ bosons. 
These divergences together with soft and collinear divergences from the initial and 
final state radiation have to be properly treated so that infrared-safe observables 
are obtained (see e.g. \cite{Catani:1996vz}). 
In this work, we have used the dipole subtraction method \cite{Catani:1996vz,Dittmaier:1999mb} 
to deal with the infrared (i.e. soft or collinear) divergences. 
More details are provided in \cite{Le:2022ppa,Dao:2023kwc,Dao:2024ffg}.

The polarized $W^+W^-$ process was first calculated in \cite{Denner:2023ehn,Dao:2023kwc} 
at NLO EW using the four-flavor scheme, i.e. the bottom-quark induced processes were excluded. 
The main reason for this choice is to exclude the top-quark contribution which occurs starting from NLO in both 
QCD and EW corrections. At NLO, the processes of $gb \to tW \to W^+W^- b$ (QCD) and $\gamma b \to tW \to W^+W^- b$ (EW)
occur. 
At NNLO, $gg \to t\bar{t} \to W^+W^- bb$ and $q\bar{q} \to t\bar{t} \to W^+W^- bb$ processes happen. 
They all produce polarized $W^+W^-$ pairs. 
These contributions are large due to the top-quark resonances. 

The definition of the inclusive $W^+W^-$ production process is
\bea
pp \to W^+W^- + N\; \text{jets} + M\; \text{photons} + X,\label{eq:WW_proc_inc}
\eea
with $N\ge 0$, $M\ge 0$. A jet here can include all partons except the top quark. 
This definition includes the default $q\bar{q}/gg \to W^+W^-$ mechanisms as well as 
the above top-quark induced ones. 
The $t\bar{t}$ contribution is completely dominant due to its large cross section at the LHC.  

In measurements, the top-quark contribution can be suppressed using jet cuts (e.g. $p^T_\text{jet-veto} = 35\gev$) and then 
subtracted in the signal region using a partly data-driven method \cite{ATLAS:2019rob}. 
The remaining contribution is the default $q\bar{q}/gg \to W^+W^-$ mechanisms, which are widely 
known in the community as the $W^+W^-$ production signal. 

For the SM prediction, the same procedure can be performed if the inclusive cross section 
of \eq{eq:WW_proc_inc} is accurately known. The top-quark background contribution must be 
separately calculated and then subtracted from the inclusive result. 
In this way, the top-quark interference contribution is part of the signal and is 
consistent with the measurement. This definition of the $W^+W^-$ signal in the five-flavor scheme (5FS)
is discussed in detail in \cite{Gehrmann:2014fva}. 

From a precision-calculation point of view, there is a good approach called the four-flavor scheme (4FS). 
In this scheme, the bottom quark is massive, and hence bottom jet is excluded in the 
inclusive cross section definition (see \eq{eq:WW_proc_inc}). 
Bottom PDF is set to zero, hence no bottom quark in the initial state. 
This 4FS definition of the $W^+W^-$ signal is free of the top-quark contribution by definition. 
The calculation is easier because the masses of the light quarks are set to zero, allowing 
for higher perturbative order computation.  
However, there is a mismatch between this definition and the measurement 
because there is still a bottom-induced contribution in the measurement result after the subtraction 
of the top-quark contribution. Moreover, the top-quark interference (i.e the signal-background interference) 
is part of the signal in the measurement result. This term is missing in the 4FS definition.  

\begin{figure}[ht!]
  \centering
  \includegraphics[width=0.8\textwidth]{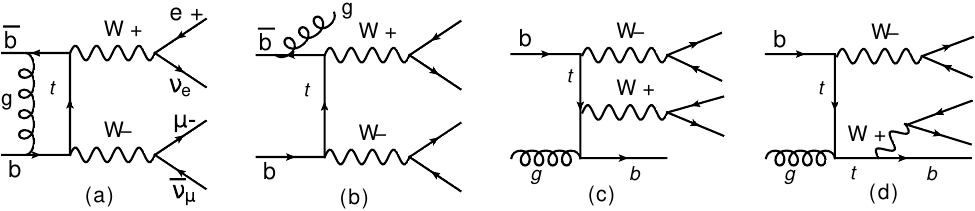}
  \caption{Various types of bottom-induced contributions to the inclusive $W^+W^-$ production. 
  (a,b,c) are of non-$tW$ origin, while (d) is of $tW$ origin. The top-quark interference (or $tW$ 
  interference) is the interference between (c) and (d).}
  \label{fig:bb_NLO_diags}
\end{figure}
This point is illustrated in \fig{fig:bb_NLO_diags}. 
The mismatch between the 4FS definition and the measurement includes contribution from 
diagrams (a,b,c) and the interference between (c) and (d).

We can account for this mismatch by calculating the bottom induced corrections separately. 
In this way we can make use of the precise calculation of the 4FS. 
This approach is well motivated because the bottom induced contribution is in most cases much smaller 
than the light-quark contribution. Moreover, as $b$ jets can be tagged in experimental analyses, 
a separate treatment of the $b$-induced contribution is a good idea.   

In this work, we include $b$ jets in the definition of the inclusive $W^+W^-$ production. 
Accordingly, the 5FS definition is used for the signal cross section. 
We will follow the same procedure as in the experimental measurement. 
Namely, the inclusive cross section including the top-quark contribution will firstly 
be calculated. Then, the on-shell top-quark contribution will be separately computed. 
The signal cross section is then obtained after the subtraction of the on-shell top-quark 
contribution. This subtraction has to be done for individual polarized cross sections. 
As our calculation is limited to NLO QCD+EW, the top-quark contribution includes only the $tW$ 
mechanism. Details of this calculation are provided in \cite{Dao:2024ffg}.  

\section{Numerical results}
\label{sect:res}
In measurements, a jet veto (applying for all partons) is usually used to reduce the top-quark backgrounds, 
as done in the ATLAS case \cite{ATLAS:2019rob}. 
The problem with this choice is that using a jet veto increases the error of the 
theoretical prediction \cite{Stewart:2011cf}. 
In the CMS measurement \cite{CMS:2020mxy}, a jet veto was used in the Sequential Cut 
but not in the Random Forest cut. Moreover, the number of $b$-tagged jets is required to be 
zero to reduce the top-quark backgrounds. In this way, jet veto is used only for the 
bottom induced contribution, thereby reducing the jet-veto error of the theoretical prediction 
as the $b$-induced cross section is small in most cases. 

In the following, we will present the results for two cut setups: YesVeto and NoVeto. 
The YesVeto setup reads
\begin{align}
        & p_{T,\ell} > 27\gev, \quad p_{T,\text{miss}} > 20\gev, \quad |\eta_\ell|<2.5, \quad m_{e\mu} > 55\gev,\crn
        & \text{jet veto (no jets with $p_{T,j}>35\gev$ and $|\eta_j|<4.5$)},
\end{align}
while the NoVeto one is identical except that the jet-veto cut is removed. 
For the YesVeto case, the $b$ jet is treated in the same way as the gluon or light-quark jets. 

\begin{table}[h!]
 \renewcommand{\arraystretch}{1.3}
\begin{center}
\setlength\tabcolsep{0.03cm}
{\fontsize{9.0}{9.0}
\begin{tabular}{|c|c|c|c|c|}\hline
  & \multicolumn{4}{c|}{YesVeto}\\
  \hline
  & $\sigma_\text{NoB}\,\text{[fb]}$ & $\sigma_\text{NoTW}\,\text{[fb]}$ & $f_\text{NoB}\,\text{[\%]}$ & $f_\text{NoTW}\,\text{[\%]}$ \\
\hline
{\fontsize{9.0}{9.0}$\text{Unpol}$} & $218.47(3)^{+2.2\%}_{-2.1\%}$ & $220.50(3)^{+2.1\%}_{-2.0\%}$ & 
$100$ & $100$ \\
\hline
{\fontsize{9.0}{9.0}$W^{+}_{L}W^{-}_{L}$} & $14.34^{+1.8\%}_{-2.6\%}$ & $15.59^{+1.2\%}_{-2.2\%}$ & 
$6.6$ & $7.1$ \\
{\fontsize{9.0}{9.0}$W^{+}_{L}W^{-}_{T}$} & $24.79^{+1.9\%}_{-2.5\%}$ & $25.31^{+1.6\%}_{-2.5\%}$ & 
$11.3$ & $11.5$ \\
{\fontsize{9.0}{9.0}$W^{+}_{T}W^{-}_{L}$} & $25.47^{+2.1\%}_{-2.5\%}$ & $25.99^{+1.8\%}_{-2.4\%}$ & 
$11.7$ & $11.8$ \\
{\fontsize{9.0}{9.0}$W^{+}_{T}W^{-}_{T}$} & $152.59(3)^{+2.2\%}_{-1.9\%}$ & $152.67(3)^{+2.2\%}_{-1.9\%}$ & 
$69.8$ & $69.2$ \\
\hline
{\fontsize{9.0}{9.0}$\text{Pol-int}$} & $1.27(4)$ & $0.93(4)$ & 
$0.6$ & $0.4$ \\
\hline
\end{tabular}
\begin{tabular}{|c|c|c|c|}\hline
\multicolumn{4}{|c|}{NoVeto}\\
  \hline
$\sigma_\text{NoB}\,\text{[fb]}$ & $\sigma_\text{NoTW}\,\text{[fb]}$ & $f_\text{NoB}\,\text{[\%]}$ & $f_\text{NoTW}\,\text{[\%]}$ \\
\hline
$327.94(4)^{+5.4\%}_{-4.2\%}$ & $334.17(4)^{+5.4\%}_{-4.1\%}$ & $100$ & $100$ \\
\hline
$18.68^{+4.1\%}_{-3.3\%}$ & $21.04(1)^{+4.0\%}_{-2.9\%}$ & $5.7$ & $6.3$ \\
$43.33^{+6.0\%}_{-4.9\%}$ & $44.86(1)^{+6.1\%}_{-4.8\%}$ & $13.2$ & $13.4$ \\
$44.22(1)^{+6.2\%}_{-4.9\%}$ & $45.77(1)^{+6.2\%}_{-4.8\%}$ & $13.5$ & $13.7$ \\
$221.43(3)^{+5.3\%}_{-4.1\%}$ & $222.80(3)^{+5.3\%}_{-4.1\%}$ & $67.5$ & $66.7$ \\
\hline
$0.28(5)$ & $-0.30(5)$ & $0.1$ & $-0.1$ \\
\hline
\end{tabular}
}
\caption{\small Results for the unpolarized and polarized DPA cross sections using the YesVeto and NoVeto cut setups. 
These results are taken from \cite{Dao:2024ffg}.}
\label{tab:xs_fr_all}
\end{center}
\end{table}
Results for the integrated cross sections are shown in \tab{tab:xs_fr_all} for the YesVeto and NoVeto cases. 
Unpolarized and polarized cross sections are provided for two cases: NoB, NoTW. 
Here NoB means the bottom-induced contribution is excluded. 
NoTW means the $b$-induced processes are included and the on-shell $tW$ contribution is subtracted. 
This is the signal cross section as measured in ATLAS and CMS. 
The polarization fractions defined as $f_i = \sigma_i/\sigma_\text{unpol}$ for the polarization $i$ 
are also provided. The polarization interference is provided in the last row. 
The scale uncertainties are shown in percent as upper and lower indices on the value of the cross section. 
These uncertainties are calculated using the direct variation method of the factorization and renormalization 
scales (seven-point method, see \cite{Dao:2024ffg} for details).
The Monte-Carlo errors are provided in the round parentheses if significant. 
We see that the $b$-induced contribution increases the LL fraction by $7.6\%$ and $10.5\%$ for the 
YesVeto and NoVeto cases, respectively. 

\begin{figure}[ht!]
  \centering
  \includegraphics[width=0.49\textwidth]{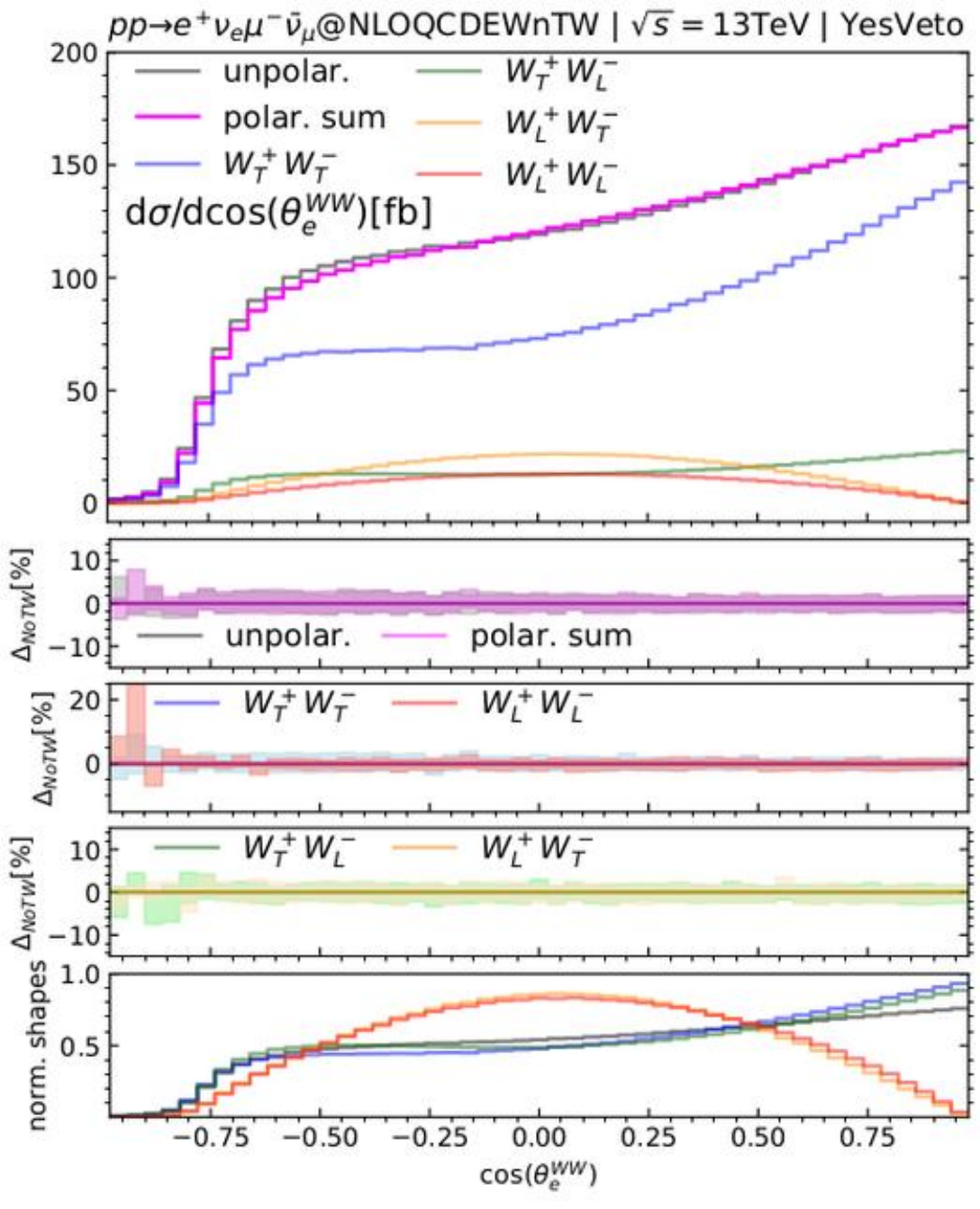}
  \includegraphics[width=0.49\textwidth]{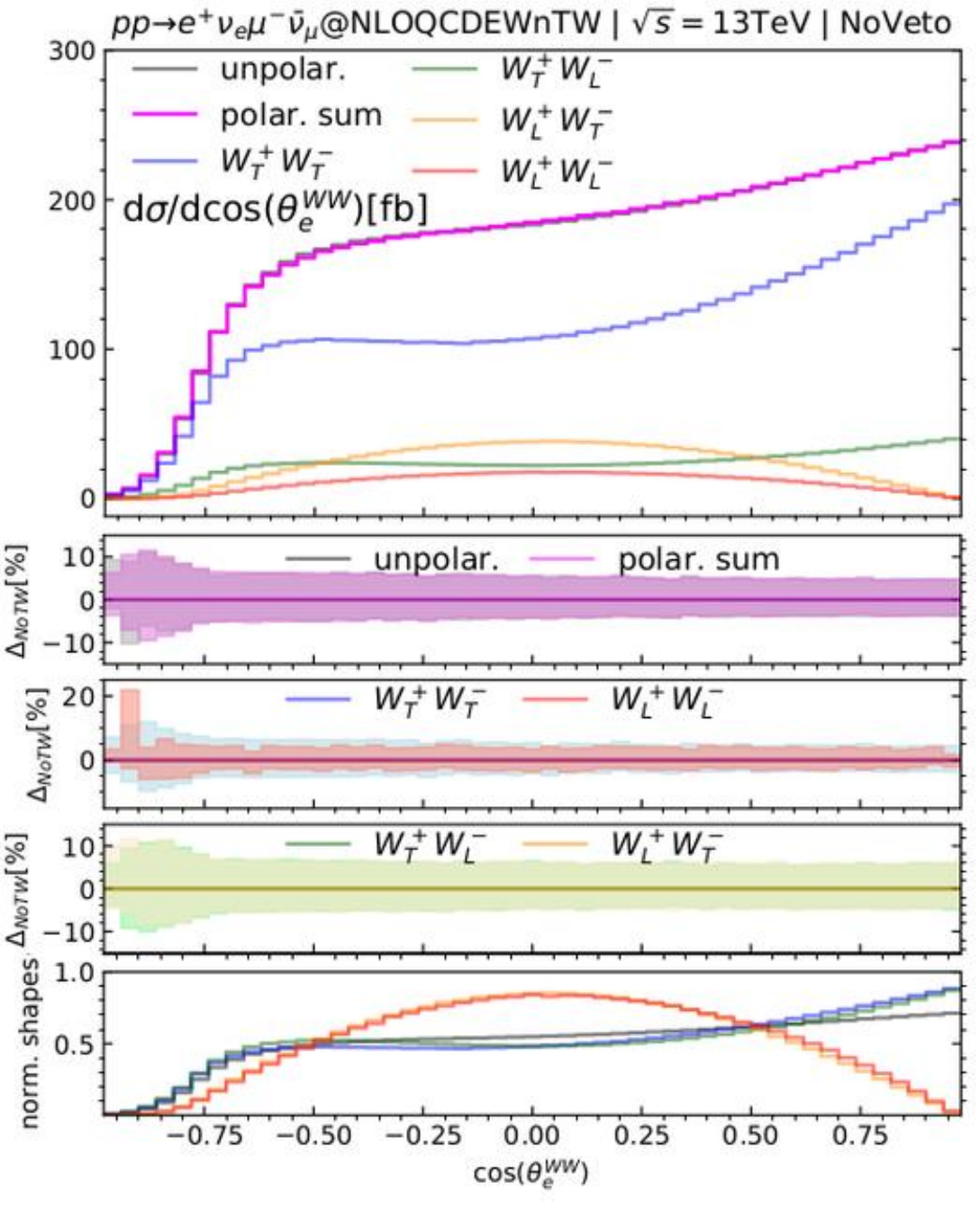}
  \caption{Differential polarized cross sections in $\cos\theta_e^{WW}$ (see text) for YesVeto and NoVeto cut setups. 
  Results are taken from \cite{Dao:2024ffg}.}
  \label{fig:diff_xs_example}
\end{figure}
Example results of differential cross sections are shown in \fig{fig:diff_xs_example}. 
The kinematic variable here is $\cos\theta_e^{WW}$, the electron polar angle calculated in the 
$WW$ center-of-mass frame. This is however not a normal polar angle. 
$\theta_e^{WW}$ is the angle between the spatial momentum $\vec{p}_e^\text{W-rest}$ determined 
in the $W^+$-rest frame and the spatial momentum $\vec{p}_{W^+}^{WW}$ determined in the 
$WW$ center-of-mass frame. 
As the Lorentz boost order is important, we have to boost the relevant momenta to the $WW$ frame first, and then 
from here boost to the $W^+$ rest frame. 
If we do not follow this order and jump directly to the $W^+$ rest frame from e.g. the LAB frame, then the 
obtained $\vec{p}_e^\text{W-rest}$ would be wrong. 

This polar angle is of particular important in the polarization measurement. 
This is because it is sensitive to the polarization of the parent gauge boson, i.e. the $W^+$ in this discussion. 
It is not sensitive to the polarization of the other gauge boson. 
This is why we get almost identical normalized shapes for the $W^+_L W^-_L$ and $W^+_L W^-_T$ cases, 
as can be seen from the bottom panels in \fig{fig:diff_xs_example}. 
Similarly, the shapes of the $W^+_T W^-_L$ and $W^+_T W^-_T$ cases are also almost identical. 
The distinct difference in shape between the $W^+_L W^-_L$ and $W^+_T W^-_L$ cases 
shows the power of this polar angle. 
Likewise, to distinguish the polarizations of the $W^-$ boson, the muon polar angle $\cos\theta_{\mu}^{WW}$ should be used.

The three small panels below the big panel show the relative scale uncertainties of the cross sections provided 
in the big panel. The uncertainties are much smaller for the YesVeto case. 
This is due to an accidental cancellation of different NLO corrections at the chosen value of the jet-veto threshold 
($35\gev$), see \cite{Stewart:2011cf}. It turns out that the cancellation is largest around this value, leading to small values of the scale uncertainties. As discussed in \cite{Dao:2024ffg}, the scale uncertainties provided here for the YesVeto case are 
likely to be underestimated. Care must therefore be taken when using these values. 
\section{Conclusions}
\label{sect:con}
In this report, we have discussed a calculation method for the inclusive polarized $W^+W^-$ production signal at the LHC in the 
five-flavor scheme. 
The method is applicable at NLO QCD+EW level. 
The key idea is to calculate the on-shell top-quark contribution separately and subtract it from the inclusive cross section. 
Example results have been presented for the integrated cross sections and a differential cross section for two cut 
setups, with and without a jet veto.
\section*{Acknowledgments}
The authors would like to thank the workshop organizers for organizing this nice event. 
This work is funded by Phenikaa University under grant number PU2023-1-A-18.

\section*{References}

\providecommand{\href}[2]{#2}\begingroup\raggedright\endgroup
\end{document}